\documentclass[prb,aps,twocolumn,showpacs]{revtex4}
\usepackage{graphicx,latexsym}
\usepackage{dcolumn}
\usepackage{amsmath,amssymb,epsf,bm}
\begin{document}
\title{Spin-Hall conductivity of a disordered 2D electron gas with
Dresselhaus spin-orbit interaction}
\author{A.~G. Mal'shukov$^1$ and K.~A. Chao$^2$}
\affiliation{$^1$Institute of
Spectroscopy, Russian Academy of Science, 142190, Troitsk, Moscow
oblast, Russia \\ $^2$ Solid State Theory Division, Department of
Physics, Lund University, S-22362 Lund, Sweden}
\date{\today}
\begin{abstract}
The spin-Hall conductivity of a disordered 2D electron gas has been
investigated for a general spin-orbit interaction. We have found that
in the diffusive regime of electron transport, the dc spin-Hall
conductivity of a homogeneous system is zero due to impurity
scattering when the spin-orbit coupling contains only the Rashba
interaction, in agreement with existing results. However, when the
Dresselhaus interaction is taken into account, the spin-Hall current
is not zero. We also considered the spin-Hall currents induced by an
inhomogeneous electric field. It is shown that a time dependent
electric charge induces a vortex of spin-Hall currents.
\end{abstract}
\maketitle

Spintronics is a fast developing area using the electron spin degrees
of freedom in electronic devices~\cite{Prinz,Wolf,Awschalom, Zutic}.
One of the most challenging goals of spintronics is to find a method
to manipulate spins by electric fields. The spin-orbit interaction
(SOI), which couples the electron momentum and spin, can serve as a
spin-charge mediator. There have been several suggestions to use the
SOI in semiconductor quantum wells (QW) to create the electron and
hole spin currents and to accumulate the spin polarization by
applying an electric field
parallel~\cite{Murakami,Sinova,Edelstein,Inoue1} or
perpendicular~\cite{MalshAC,Governale} to the QW. The spin current
induced by the parallel electric field and flowing perpendicular to
it has been named the spin-Hall effect (see also~\cite{Hirsch}).
Since the prediction of this effect by Murakami
{\it et. al.}~\cite{Murakami} and Sinova {\it et. al.}~\cite{Sinova},
there have been much discussions concerning the effect of nonmagnetic
impurity scattering on the spin-Hall conductivity in systems with
Rashba spin-orbit coupling. Some groups predicted that the impurity
scattering should suppress the spin-Hall effect induced by a
homogeneous and static electric
field~\cite{Inoue+,Mischenko,Dimitrova, Raimondi}, even if the mean
scattering time $\tau$ is much longer than $1/\Delta$, where $\Delta$
is the spin-orbit splitting of the electron energy (we set
$\hbar$=1). This result was confirmed by an analysis of the sum rules
in Ref.~\cite{Rashba}. Yet some other groups came to different
conclusions~\cite{Burkov,Chelaev,Nomura}.

In the present we use the diffusion approximation to derive an
expression of the spin-Hall conductivity for a general SOI including
both Rashba and Dresselhaus terms. For pure Rashba SOI, as well as
for linear Dresselhaus interaction, we found that the dc spin-Hall
conductivity of the homogeneous system becomes zero even for a weak
disorder scattering, confirming thus the results of
Ref.~[\onlinecite{Inoue+,Mischenko,Dimitrova,Raimondi, Rashba}].
On the other hand, when the cubic terms of Dresselhaus SOI is
included, a finite spin current is produced. In order to study the
effect of a spatially inhomogeneous electric field, our analysis
keeps finite frequency $\Omega$ and wavenumber $\bm{Q}$ of the
electric field. We found that for $\Omega$$\ll$$DQ^2$, where $D$ is
the electron diffusion constant, the flow of the spin-Hall currents
is dominated by the screening effects. Similar to formation of an
electron screening cloud around an external charge, the spin-Hall
currents form a vortex.

We consider a typical III-V semiconductor QW with only the lowest
subband occupied. The spin-orbit coupling of conduction electrons
has the form
\begin{equation}\label{hso}
H_{so} = \bm{h}_{\bm{k}}\cdot\bm{\sigma} \, ,
\end{equation}
where $\bm{\sigma}$$\equiv$$(\sigma^x,\sigma^y,\sigma^z)$ is the
Pauli matrix vector, and $\bm{h}_{\bm{k}}$ a function of the
two-dimensional wave-vector $\bm{k}$. In general, $\bm{h}_{\bm{k}}$
contains both the Dresselhaus and the Rashba terms. The former exists
also in bulk crystals~\cite{dress}, while the latter appears only in
asymmetric QWs~\cite{rash}. For a QW grown along the [001] direction,
which is set as the $z$ axis, the Dresselhaus SOI is given
by~\cite{calc}
\begin{eqnarray}\label{hD}
h^x_{\bm{k}} &=& \beta k_x (k^2_{y} - a^2) \, , \\ \nonumber
h^y_{\bm{k}} &=& -\beta k_y (k^2_{x} - a^2) \, ,
\end{eqnarray}
where the parameter $a^2$ is the average of the operator
--$(\partial/\partial z)^2$ with respect to the lowest subband wave
function. The Dresselhaus SOI in (\ref{hD}) contains terms both
linear and cubic in $\bm{k}$. Usually, in heavily doped QWs, for
electrons at the Fermi energy both terms are of the same order of
magnitude~\cite{Jusserand}. The Rashba interaction has the
form~\cite{rash}
\begin{equation}\label{hR}
h^x_{\bm{k}}=\alpha k_y \,\,\,\,\, ; \,\,\,\,\,
h^y_{\bm{k}}= -\alpha k_x \, .
\end{equation}

Let us apply an electric field along the $x$ axis, and express it
as the gradient of a scalar electric potential
$\bm{E}$=--$\bm{\nabla}V$. This gauge is more convenient for studying
the case of finite wave-numbers $\bm{Q}$ in the Fourier expansion of
$\bm{E}$. The one-particle spin-current operator is
$J^i_j$=($\sigma^iv^j$+$v^j\sigma^i$)/4, where the particle velocity
is
\begin{equation}\label{v}
v^i = \frac{k^i}{m*} + \frac{\partial}{\partial
k^i}(\bm{h}_{\bm{k}}\cdot\bm{\sigma}) \, .
\end{equation}
This definition has to be used with cautious, since the spin current
is not conserving in systems with SOI, as discussed in
Ref.~\onlinecite{Rashba1}. We are interested in calculating the spin
current polarized in $z$ direction and flowing in $y$ direction.
Since $\bm{h}_{\bm{k}}$ in (\ref{hR}) and (\ref{hD}) has no $z$
components the spin-current operator is $J^z_y$=$\sigma^zk^y/(2m^*)$.
We will calculate the corresponding spin-Hall current within the
standard linear response theory~\cite{agd} and denote it as $J$. So,
the initial expression for $J$ is
\begin{eqnarray}\label{response}
J &=& -ie\Omega\sum_{\bm{k},\bm{k'}} \int \frac{d\omega}{2\pi}
\frac{\partial n_F(\omega)}{\partial\omega}\langle
Tr[G^a(\bm{k}_-,\bm{k'}_-,\omega) \times \nonumber \\
&& \times J^z_y G^r(\bm{k}^{\prime}_+,\bm{k}_+,\omega+\Omega)]
\rangle \, V(\Omega,\bm{Q}) \, ,
\end{eqnarray}
where $\bm{k}_{\pm}$=$\bm{k}$$\pm$$\bm{Q}/2$, and $n_F(\omega)$ is
the Fermi distribution function. In (\ref{response}) the trace runs
through the spin variables, and the angular brackets denote the
average over the random distribution of impurities. The terms
containing the products of the form $G^aG^a$ and $G^rG^r$ are
neglected since their contribution to the spin-Hall current is
small~\cite{Raimondi}. For simplicity we assume that in the vicinity
of the Fermi energy $E_F$, the amplitude of impurity elastic
scattering is isotropic and momentum-independent. In the
quasiclassical approximation, when $E_F\tau$$\gg$1, the average of
the product of the retarded and advanced Green functions $G^r$ and
$G^a$ can be calculated perturbatively. If we ignore weak
localization effects, the perturbation expansion of (\ref{response})
consists of the so called ladder diagrams~\cite{agd,alt}. For small
$\Omega$ and $Q$ these diagrams describe the particle and spin
diffusion processes. The spin diffusion also includes the
D'yakonov-Perel spin relaxation~\cite{dp}. Therefore, the spin-Hall
current (\ref{response}) is determined by the combination of spin and
particle diffusion propagators.

To calculate and to combine these propagators for arbitrary
$\bm{h_{\bm{k}}}$, we will follow the formalism of
Ref.~[\onlinecite{MalshHall,MalshWL}]. In (\ref{response}) the
spin-current vertex $J^{z}_y$ is coupled to the spin independent
potential $V$. Such a spin-charge coupling has two channels. In the
first channel, $J^{z}_y$ and $V$ are coupled via the spin-independent
particle diffusion propagator. This contribution to the spin-Hall
current is denoted as $J_1$. For $\Omega$$\ll$$1/\tau$ and
$v_FQ$$\ll$$1/\tau$, where $v_F$ is the Fermi velocity, from
(\ref{response}) we obtain
\begin{equation}\label{J1}
J_1 = i\frac{e\Omega}{2\pi}\Psi\,D(\Omega,\bm{Q})\,
V(\Omega,\bm{Q}) \, ,
\end{equation}
where $D(\Omega,Q)$=$[\tau(-i\Omega+DQ^2)]^{-1}$ is the particle
diffusion propagator~\cite{alt}. The vertex $\Psi$ is
\begin{equation}\label{psi0}
\Psi = \sum_k Tr[G^r(\bm{k}_+,E_F + \Omega)
G^a(\bm{k}_-,E_F)J^z_y] \, ,
\end{equation}
where $G^{r,a}(\bm{k},E)$ are the Green functions averaged over
random impurity positions.

The second coupling channel is more complicated. The spin current
couples first to the spin diffusion-relaxation propagator, which
couples to $V$ via the mixing of charge and spin diffusion processes.
The mixing of these diffusion processes was pointed out explicitly by
Burkov {\it at. al.}~\cite{Burkov}. The spin-Hall current due to this
channel is denoted as $J_2$, and is obtained as
\begin{equation}\label{J2}
J_2 = i\frac{e\Omega}{2\pi} \Psi^l\,D^{lj}(\Omega,\bm{Q}) \,
M^j\,D(\Omega,Q ) \, V(\Omega,\bm{Q}) \, ,
\end{equation}
with the vertices
\begin{equation}\label{psi}
\Psi^l = \sum_k Tr[G^r(\bm{k}_+,E_F + \Omega) \sigma^l
G^a(\bm{k}_-,E_F)J^z_y] \, .
\end{equation}
In (\ref{J2}) the superscripts $l$ and $j$ are summed over $x$, $y$
and $z$. The spin diffusion-relaxation propagator
$D^{ij}(\Omega,\bm{Q})$ describes diffusion and relaxation of a spin
density packet. Therefore, this propagator satisfies the spin
diffusion equation for spins polarized in the $j$-direction when a
source creates spins polarized in the $i$-direction. $M^j$ is the
spin-charge mixing, defined as
\begin{eqnarray}\label{Mi}
M^j &=& \frac{1}{4\pi\tau N_0}\sum_k
Tr[G^r(\bm{k}_+,E_F + \Omega) \times \nonumber \\
&& \times G^a(\bm{k}_-,E_F)\sigma^j] \, ,
\end{eqnarray}
where $N_0$=$m^*/(2\pi)$ is the 2D density of states. $M^j$ makes the
diffusion of spins polarized in $j$-direction dependent on the charge
density distribution~\cite{Burkov,Mischenko}. This spin-charge
coupling is weak and is proportional to the small parameter
$h_{\bm{k}}/E_F$. Therefore, in (\ref{J2}) we keep only the terms
linear in $M^j$. It should be noticed~\cite{Edelstein} that $J_2$ is
closely related to the electric field induced accumulation of the
in-plane polarized spin density $S^l$. For example, it can be shown
that $J_2$=$\Psi^lS^l/2\pi\tau N_0$.

After averaging over the impurity positions, the retarded and
advanced Green functions are obtained as
\begin{equation}\label{Green}
G^r(\bm{k},E) = [G^{a}(\bm{k},E)]^{\dag} =
(E - E_{\bm{k}} - \bm{h}_{\bm{k}}\cdot\bm{\sigma} +
i\Gamma)^{-1} \, ,
\end{equation}
where $\Gamma$=$1/(2\tau)$ and $E_{\bm{k}}$=$k^2/(2m^*)$. For the
case of short-range impurities and the constant density of states
near $E_F$, the scattering rate $\Gamma$ is independent of
momentum~\cite{agd}. Using (\ref{Green}), for small $\Omega$ and
$\bm{Q}$, one gets from (\ref{psi0}), (\ref{psi}) and (\ref{Mi})
\begin{eqnarray} \label{psi2}
\Psi &=& \frac{i\pi N_0}{2\Gamma}\, \epsilon^{lmz}\,Q^n \,
\overline{(\nabla^n_{\bm{k}} h^l_{\bm{k}}) h^m_{\bm{k}} v^y
Z_{\bm{k}}} \, ,\nonumber \\
\Psi^l &=& -\pi N_0\, \epsilon^{lmz} \, \overline{v^y
h^m_{\bm{k}}Z_{\bm{k}}} \, , \nonumber \\
M^j &=& \frac{i}{2\Gamma} Q^m \, \overline{(\nabla^m_{\bm{k}}
n^j_{\bm{k}})h^3_{\bm{k}}Z_{\bm{k}}} \, ,
\end{eqnarray}
where $Z_{\bm{k}}$=($\Gamma^2$+$h^2_{\bm{k}}$)$^{-1}$ and
${\bm{n}}_{\bm{k}}$$\equiv$${\bm{h}}_{\bm{k}}/h_{\bm{k}}$. The
over-line in (\ref{psi2}) denotes the average over directions of
$\bm{k}$ which has the magnitude $k$=$k_F$. In (\ref{psi2})
$\epsilon^{lmz}$ is the antisymmetric tensor with $\epsilon^{xyz}$=1,
and all doubly repeated superscripts should be summed over $x$, $y$
and $z$.

$D^{ij}(\Omega,\bm{Q})$ satisfies the spin diffusion
equation~\cite{Mischenko,MalshDiff}. For $Q v_F$$\ll$$h_{\bm{k}_F}$
we can neglect in this equation the diffusion and spin precession
terms which are proportional to the gradient of the spin propagator.
We then have
\begin{equation}\label{diffusion}
-i\Omega D^{mj}(\Omega,\bm{Q}) = 2\Gamma\delta^{mj} -
\Gamma^{ml}D^{lj}(\Omega,\bm{Q}) \, ,
\end{equation}
where $\Gamma^{ml}$ is the spin relaxation matrix element. At low
frequency the relaxation term dominates and so
$D^{mj}(\Omega,\bm{Q})$ is simply given by the inverse of
$\Gamma^{ml}$, and
\begin{equation}\label{gamma}
\Gamma^{ml} = 2\Gamma\, \overline{[\delta^{ml} h^2_{\bm{k}}
- h^m_{\bm{k}} h^l_{\bm{k}}]Z_{\bm{k}}} \, .
\end{equation}
This equation differs by a factor $\Gamma^2 Z_{\bm{k}}$ from the
standard definition of the spin relaxation matrix, for example, in
Ref.~[\onlinecite{MalshHall}]. This factor is not unity because we
consider the situation that the spin splitting $\Delta$=$2h_{\bm{k}}$
can be comparable to the electron elastic scattering rate $2\Gamma$.

Let us first consider the case of Rashba SOI (\ref{hR}). We then set
$Q^y$=0 and $E$=--$iQ^x V$ to calculate $\Psi$, $\Psi^y$ and $M^y$
from (\ref{psi2}). In this case both the spin relaxation matrix and
the spin diffusion-relaxation propagator are diagonal. Substituting
the so calculated $\Psi$, $\Psi^y$, $M^y$ and $D^{yy}$ into
(\ref{J1}) and (\ref{J2}), the currents $J_1$ and $J_2$ are obtained
as
\begin{equation}\label{currentR}
J_1 = -J_2 = E\frac{e}{8\pi}\frac{\Delta^2}{4\Gamma^2 +
\Delta^2}\frac{\Omega}{\Omega+iDQ^2}\, ,
\end{equation}
where $\Delta$=$2\alpha k_F$. Hence, the total current $J_1$+$J_2$
vanishes even for small impurity scattering rate
$\Gamma$$\ll$$\Delta$, in agreement with the existing
results~\cite{Inoue+,Mischenko,Rashba,Dimitrova,Raimondi}. We should
mention that in deriving this result for $\Omega$$\ll$$\Gamma^{yy}$,
in the denominator (--$i\Omega$+$\Gamma^{yy}$) of the spin
diffusion-relaxation propagator the frequency term has been removed.
If we retain $\Omega$, $J_1$ and $J_2$ will cancel each other not
exactly, but the accuracy~\cite{Mischenko} is up to
$\Omega/\Gamma^{yy}$. As was pointed out by Mishchenko {\it et.
al.}~\cite{Mischenko}, near the sample boundaries $J_2$ can also
differ from $J_1$ because of the rapid spatial variation of the spin
diffusion propagator. We have ignored this effect by neglecting the
gradient terms in the diffusion equation (\ref{diffusion}). If
necessary, in our approach we can consider the boundary problem by
substituting into (\ref{J2}) the complete solution
$D^{mj}(\Omega,\bm{Q})$ of the spin diffusion
equation.~\cite{MalshDiff} Our main goal is, however, to show that
the spin current is not zero in the bulk of the sample when the
Dresselhaus SOI is taken into account. In this case the total spin
accumulation near the sample edge will be determined by a direct
inflow of the spin polarization from the bulk.

Let us assume that the SOI contains only the Dresselhaus interaction
(\ref{hD}), which has terms both linear and cubic in $\bm{k}$. When
the cubic interaction is ignored, there is no spin-Hall effect
because the linear Dresselhaus SOI can be obtained from the Rashba
SOI via a unitary transformation of the spin operators~\cite{Rashba}.
For the complete Dresselhaus interaction (\ref{hD}), following
(\ref{J1}), (\ref{J2}) and (\ref{psi2})--(\ref{gamma}), the
calculation of the spin-Hall current is straightforward. We obtain
the total spin current $J$=$J_1$+$J_2$ as
\begin{equation}\label{jtot}
J = E \,\sigma_{sH}\, \frac{\Omega}{\Omega+iDQ^2} \, ,
\end{equation}
where $\sigma_{sH}$ is the DC spin-Hall conductivity at
$Q$$\rightarrow$0. The calculated $\sigma_{sH}/(e/16\pi)$ is plotted
in Fig.~1 as a function of $a/k_F$, for three values of
$\Gamma^2/\beta^2k_F^6$=10$^{-4}$, 10$^{-3}$, and 10$^{-1}$. The
ratio $a/k_F$ is a measure of relative strength of the linear to
cubic terms in (\ref{hD}). As expected, the $\sigma_{sH}$ vanishes
for large $a$. It is important to notice the singular behavior at
small $\Gamma$ of $\sigma_{sH}$ in the vicinity of
$a/k_F$=1/$\sqrt{2}$ and $a/k_F$=0. The singularities appear because
at these points the spin-orbit splitting $2h_{\bm{k}}$ vanishes for
certain $\bm{k}$ directions. As a result, in such angular integrals
$Z_{\bm{k}}^{-1}$$\rightarrow$$\infty$ when the elastic scattering
rate $\Gamma$$\rightarrow$0. It is also interesting to notice that in
the range 0$<$$a/k_F$$<$$1/\sqrt{2}$, as $\Gamma$$\rightarrow$0 the
spin-Hall conductivity has a plateau shape with the universal value
of $\sigma_{sH}$=$3e/8\pi$. This plateau and the sharp change of sign
at $a/k_F$=1/$\sqrt{2}$ can be useful in device applications.
\begin{figure}[tbp]
\includegraphics[angle=270,width=7.5cm]{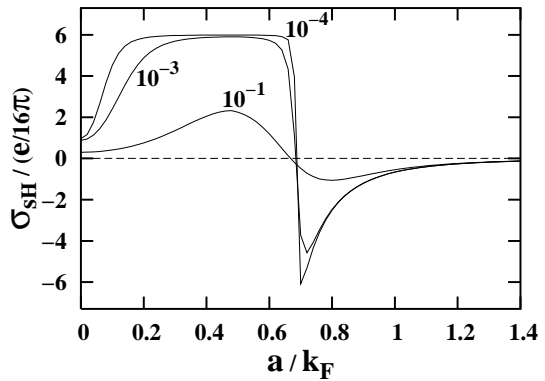}
\caption{Spin-Hall conductivity as a function of $a/k_F$ for
$\Gamma^2/\beta^2 k^{6}_{F}$=10$^{-4}$, 10$^{-3}$, and 10$^{-1}$.}
\label{fig1}
\end{figure}

We would like to elaborate the non-analytic behavior of (\ref{jtot})
when both $\Omega$ and $Q$ approach zero, a consequence of the
diffusion denominator in $J$. When $Q$$\rightarrow$0 first,
(\ref{jtot}) gives the DC flow of the spin-Hall current induced by
the spatially homogeneous electric field. At the opposite regime
$DQ^2$$\gg$$\Omega$, we neglect the $\Omega$ in the denominator and
rewrite (\ref{jtot}) in a coordinate independent form as
\begin{equation}\label{currentvec}
J_l =\frac{i\Omega\sigma_{sH}}{DQ^2}\, \epsilon^{ljz}E^j \, .
\end{equation}
Here $J_l$ is the $z$--polarized spin current flowing along the $l$
axis. To arrive at (\ref{currentvec}) we have assumed that $DQ^2$ is
much less than the spin relaxation rate. Otherwise, the term $DQ^2$
should be added to (\ref{diffusion}).

(\ref{currentvec}) yields the hydrodynamics of the spin-Hall current
flow. Since $\bm{E}$=--$i\bm{Q}V$ is a longitudinal field, we have
\begin{equation}\label{hydro}
\bm{\nabla}\cdot\bm{J} = 0 \,\,\,\,\, ; \,\,\,\,\,
(\bm{\nabla}\times\bm{J})_z = \frac{\sigma_{sH}}{D} \frac{\partial
V}{\partial t} \, .
\end{equation}
The first equation indicates that the spin current is conserving. The
second equation tells us that in each spatial point the flux is
perpendicular to the local electric field, similar to the spin-Hall
effect in a homogeneous field. In the field of spherically symmetric
potential a circular vortex flow of the spin current is thus induced
around a central charge. The physics of this effect is similar to the
screening of scalar potential by electric charges. To clarify this
analogy, let us introduce the conjugate current $\tilde{J}_x$=$J_y$
and $\tilde{J}_y$=--$J_x$, as well as the vortex "charge" density
$\rho$ defined by the continuity equation
\begin{equation}\label{charge}
e \nabla \tilde{\bm{J}} = \frac{\partial \rho}{\partial t} \, .
\end{equation}
We can then rewrite the second equation in (\ref{hydro}) as
\begin{equation}\label{div}
\rho=e \frac{\sigma_{sH}}{D}V \, ,
\end{equation}
which has the same form as the equation for the electrostatic
screening of the scalar potential $V$, with $e\sigma_{sH}/D$ playing
the role of the inverse screening length.

It should be noticed that because of the above mentioned close
relationship between the spin-Hall effect and the accumulation of
in-plane spin polarization, the latter will also appear as a
screening cloud around the external charge. The in-plane
polarization, in its turn, can give rise to a z-polarized component
via the spin precession term of the diffusion
equation~\cite{MalshDiff}. This precession is proportional to
$v_FQ/\Gamma$, which is small in the diffusion approximation and was
neglected in (\ref{hydro}). Consequently, the spin-Hall current
turns out to be conserved, as one can expect in the absence of the
relaxation of z-polarization. On the other hand, in the near vicinity
of the vortex core, the precession term becomes more important
because of the larger gradient of the electric field. Hence, the
accumulation of the z-polarized spin density will be expected in the
region of the core. The detailed analysis of this phenomenon is
outside the scope of the present paper. It is worthwhile to notice
that the core has a macroscopic size about $\hbar v_F/\Delta$, which
is of the order microns. Therefore, the spin accumulation in the
vortex core can be observed by, for example, the method of Faraday
rotation~\cite{Kikkawa}.

In conclusion, within the quasiclassical perturbation theory we have
shown that, in agreement with existing results, impurity scattering
reduces the DC spin-Hall current to zero if the SOI is due to the
Rashba interaction. On the other hand, the spin-Hall current remains
finite for the Dresselhaus SOI. Nevertheless, this current becomes
zero if it is induced by a {\it spatially} varying DC electric field.
The field must be time dependent in order to produce a finite effect.
In this case the spin-current flow in the field of a scalar potential
has the form of a vortex. The physics of this phenomenon is formally
equivalent to the screening of external electric potential by
electrons.

We acknowledge useful discussions with E. I. Rashba and E. G.
Mishchenko. This work was supported by the Swedish Royal Academy
of Science, the Russian Academy of Sciences and the RFBR grant No
030217452.

\end{document}